\input{aipcheck}

\documentclass[
    ,final            % use final for the camera ready runs
  ]
  {aipproc}

\layoutstyle{6x9}

\def\ni{\noindent}

\def\about{$\sim$}
\def\arcsec{$\,^{\prime\prime}$~}
\def\arcmin{$\,^\prime$}

\def\erg/cm2sec{ergs~cm$^{-2}$~s$^{-1}$}  
\def\ergcm2{ergs~cm$^{-2}$}  
  
\def\mdot{$\dot{m}$~}  
\def\Edot{$\dot{E}$~}  
\def\X{$\times$}

\def\Lx{L$_x$~}

\def\Msun{$M_\odot$}

\def\pc3{pc$^{-3}$~}
\def\cm3{cm$^{-3}$~}
\def\rc{r$_c$~}
\def\chandra{{\it Chandra}}
\newcommand{\lsim }{{\lower0.8ex\hbox{$\buildrel <\over\sim$}}}
\newcommand{\gsim }{{\lower0.8ex\hbox{$\buildrel >\over\sim$}}}
\newcommand{\atoms}{\ifmmode{\rm ~atoms~cm^{-2}} \else ~atoms cm$^{-2}$\fi}
\newcommand{\cmsq}{\ifmmode{\rm ~cm^{-2}} \else cm$^{-2}$\fi}
\newcommand{\nh}{{\ifmmode{\rm N_{H}} \else N$_{H}$\fi}~}
\newcommand{\fx}{\ifmmode L_x \else $~f_x$\fi}
\newcommand{\fxfopt}{\ifmmode \frac{f_x}{f_{opt}} \else $\frac{f_x}{f_{opt}}$\fi
}
\newcommand{\logfx}{\ifmmode{\rm log}~f_x \else log$~f_x$\fi}
\newcommand{\logfxfopt}{\ifmmode{\rm log}\,(\frac{f_x}{f_{opt}}) \else
${\rm log}\,(\frac{f_x}{f_{opt}})$ \fi}
\newcommand{\lopt}{\ifmmode L_{opt} \else $~L_{opt}$\fi}
\newcommand{\loglopt}{\ifmmode{\rm log}~L_{opt} \else log$~L_{opt}$\fi}
\newcommand{\lx}{\ifmmode L_x \else $~L_x$\fi}
\newcommand{\loglx}{\ifmmode{\rm log}~L_x \else log$~L_x$\fi}
\newcommand{\fcgs}{\ifmmode {\rm erg~cm}^{-2}~{\rm s}^{-1}\else
erg~cm$^{-2}$~s$^{-1}$\fi} 
\newcommand{\lcgs}{\ifmmode erg~~s^{-1}\else erg~s$^{-1}$\fi}
\newcommand{\flamcgs}{\ifmmode erg~cm^{-2}~s^{-1}~Hz^{-1} \else erg~cm$^{-2}$~s$
^{-1}~$\AA$^{-1}$\fi}
\newcommand{\fnucgs}{\ifmmode erg~cm^{-2}~s^{-1}~Hz^{-1}\else erg~cm$^{-2}$~s$^{
-1}$~Hz$^{-1}$\fi}
\newcommand{\lnucgs}{\ifmmode erg~s^{-1}~Hz^{-1}\else erg~s$^{-1}$~Hz$^{-1}$\fi}
\newcommand{\kms}{\ifmmode~{\rm km~s}^{-1}\else ~km~s$^{-1}~$\fi}
\newcommand{\mone}{\ifmmode ^{-1}\else$^{-1}$\fi}
\newcommand{\mtwo}{\ifmmode ^{-2}\else$^{-2}$\fi}
\newcommand{\degs}{\ifmmode ^{\circ}\else$^{\circ}$\fi}
\newcommand{\mv}{\ifmmode {m_{V}}\else${m_{V}}$\fi}
\newcommand{\Mv}{\ifmmode {M_{V}}\else${M_{V}}$\fi}
\newcommand{\msun}{\ifmmode {M_{\odot}}\else${M_{\odot}}$\fi}
\newcommand{\rsun}{\ifmmode {R_{\odot}}\else${R_{\odot}}$\fi}
\newcommand{\lsun}{\ifmmode {L_{\odot}}\else${L_{\odot}}$\fi}

\begin{document}

\title{Interacting X-ray Binaries in Globular Clusters: 
47Tuc vs. NGC 6397}


%\classification{<Replace this text with PACS numbers; choose from this list:
%                \texttt{http://www.aip..org/pacs/index.html}>}


\author{Jonathan E. Grindlay}{
  address={Harvard Observatory, 60 Garden St., Cambridge, MA 02138}
}

\begin{abstract}
Our deep \chandra~ exposures of 47Tuc and moderate 
exposures of NGC 6397 
reveal a wealth of new phenomena for interacting X-ray 
binaries (IXBs) in globular clusters. In 
this (late) Review, updated since the conference, I summarize recent 
and ongoing analysis of the millisecond pulsars, the compact binaries
containing white dwarfs and neutron stars, and the 
chromospherically active binaries in both globular clusters. 
Spectral variability analysis enables new insights into 
source properties and evolutionary history. These 
binary populations, now so ``easily'' visible, are large enough that 
their properties and spatial distributions reveal new hints of 
compact object formation and binary interactions 
with their parent cluster. Neutron stars appear overabundant, 
relative to white dwarfs, in 47Tuc vs. NGC 6397. 
The IXBs containing neutron stars (i.e., MSPs and qLMXBs),  
as the most massive and ancient compact binary sample, may trace the 
protocluster disk in 47Tuc, whereas compact binaries may have been  
ejected preferentially along the cluster rotation 
equator during the recent core collapse in NGC 6397.

\end{abstract}

\maketitle

%%%%%%%%%%%%%%%%%%%%%%%%%%%%%%%%%%%%%%%%%%%%
%% MAINMATTER
%%%%%%%%%%%%%%%%%%%%%%%%%%%%%%%%%%%%%%%%%%%%

\section{Introduction}
Globular clusters continue to delight interacting binary afficionados. 
Whereas only \about30y ago binaries were virtually unknown in 
globulars and intermediate mass black holes (IMBHs) were thought 
to be (e.g. Bahcall and Ostriker 1975) the objects responsible 
for the population of luminous X-ray sources discovered in 4 
globular clusters, compact binaries are now known (e.g. Hut 
et al 1992) to ``rule'' the dynamics of globular clusters. 
Still, it is only now becoming clear with the sharp X-ray eye of 
\chandra ~just how numerous and interactive compact binaries are 
when in globular clusters. Not only are there the 
accreting white dwarfs, or cataclysmic variables (CVs), 
as the dominant population of low luminosity 
compact X-ray binaries, as well as the (much) smaller population of 
quiescent low mass X-ray binaries (qLMXBs) -- both originally 
suspected from the original {\it Einstein} survey 
(Hertz and Grindlay 1983) -- but also the significant population of 
primordial binaries containing main sequence (and sub-giant) stars 
detected by their coronal emission as ``active binaries'' (ABs).  
More unexpected was the  discovery with the first \chandra 
observation of 47Tuc (Grindlay et al 2001a; herafter GHE01a) that 
millisecond pulsars (MSPs) 
are ``easily'' detected in globulars by their thermal 
as well as (later recognized for at least 3 of the 
19 MSPs with precise locations in47Tuc) non-thermal pulsar wind shock 
emission. Magnetospheric emission, as from more luminous 
pulsars, is not generally detected.  Other surprises have
come from \chandra ~and XMM observations of many other globular 
clusters (see Verbunt and Lewin 2005 and Verbunt, these 
proceedings, for recent reviews).  

Here I review the compact 
binary populations in two particularly interesting, and contrasting, 
globular clusters, NGC 104 (hereafter 47Tuc) and NGC 6397. 
I present, or at least touch on, results that suggest the entire suite 
of low luminosity X-ray sources in globulars -- CVs, ABs, MSPs 
and qLMXBs -- are truely interacting X-ray binaries (IXBs), with 
evidence from each type of source for interactions of the IXB 
with other cluster members or with the cluster as a whole. Details 
are given in followup papers.

\section{X-ray Overview of 47Tuc}
47Tuc remains the ``model'' globular cluster for a wide range of 
studies, particularly for IXBs. The very low absorption 
column \nh = 1.3 \X 10$^{20}$ \cmsq allows maximum 
sensitivity for very soft sources 
like the thermal emission from the polar caps of MSPs or Polars, and 
the well-measured cluster dynamics provide a framework for 
considering sources within the cluster context. Our initial (Mar. 2000) 
\chandra~ observation (70 ksec, ACIS-I) was presented by GHE01a, and 
an initial detailed study of the MSPs by Grindlay et al (2002; 
hereafter GCH02).  
The rich harvest of IXBs (108 within the central 2.5\arcmin 
\X 2.5\arcmin) merited a deep (4 \X 65ksec; Oct. 2002) 
followup with ACIS-S. 
Initial results and the overall source catalog are presented 
by Heinke et al (2005a; hereafter HGE05a, and see Heinke, these 
Proceedings).  

\subsection{New limits for a central IMBH}
\begin{figure}
\includegraphics[height=.3\textheight,clip]{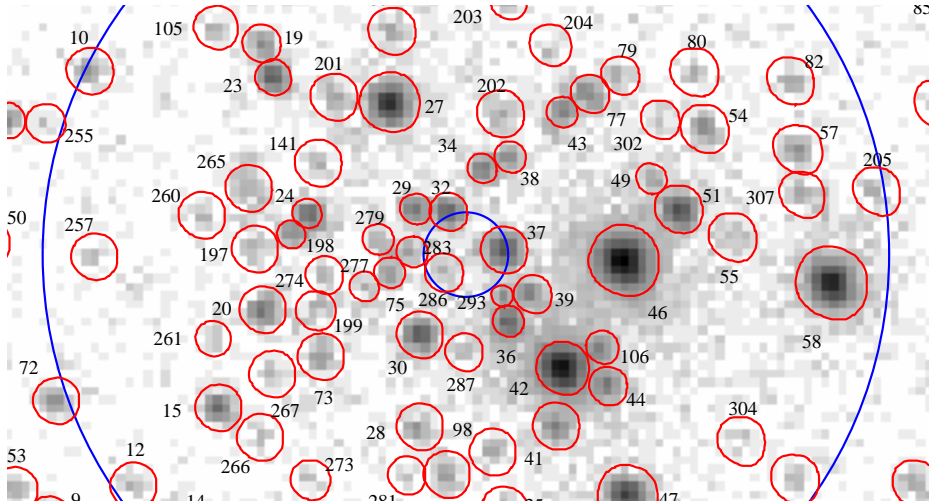}
\caption{\chandra~ image of central region of 47Tuc with circles 
marking optical core radius (24\arcsec) and estimated uncertainty 
in cluster center position (2.5\arcsec). Source W37 
is now identified as a 
qLMXB, leaving only W286 as a contender for a central IMBH.}
\end{figure}

One of the new results enabled by the first \chandra~ observation of 
47Tuc was the first restrictive X-ray limit for the presence of an 
IMBH in a globular cluster. Thanks to the detection of hot 
gas in 47Tuc (the {\it only} globular in which gas has 
been detected) by the variable dispersion measure (DM) of its 
MSPs (Freire et al 2001) and the 0.5\arcsec spatial resolution 
of \chandra, the brightest \chandra~ source with position 
consistent with being at the precise cluster center allowed a 
IMBH mass limit of 470\Msun~ to be derived assuming 
Bondi-Hoyle accretion with a (low) efficiency 
($\epsilon \sim10^{-4}$) advection flow (GHE01a). The limiting 
source within the \about2.5\arcsec uncertainty region for the 
cluster center was source W37, for which the source luminosity 
was \Lx \about1 \X 10$^{31}$ \lcgs. This source has since 
been identified as a probable qLMXB 
(Heinke et al 2005b; hereafter HGE05b), leaving only source W286 
as a contender (W32 is just out of the error circle, but 
since a IMBH could ``wander'', it is also a possible 
candidate and with \Lx similar to W37 would yield a similar 
mass limit; see Fig. 1). The \about10\X lower 
luminosity of W286 reduces by a factor of \about3  
the upper limit for IMBH mass (GHE01a),  \\ 

M(IMBH) \lsim ~~ [\Lx(0.5-2.5keV)/(4.5 \X 10$^{25} \epsilon_{-4}$ 
T$_{100keV}$)]$^{0.5}$ ~ \about150\Msun, \\

\ni
where \Lx is evaluated in the 0.5-2.5 keV band 
and is 1.2 \X 10$^{30}$ \lcgs for W286 (HGE05a), $\epsilon_{-4}$ 
is the advection-accretion efficiency in units of 10$^{-4}$, 
and T$_{100keV}$ is the advection-accretion temperature in 
units of 100 keV. The normalization again assumes the ISM in 
47Tuc has density 0.1cm$^{-3}$ as suggested by the DM 
variations of the MSPs. If the MSP winds have evacuated a 
bubble in the central core of 47Tuc (note that MSP-W = 
W29, for which the MSP wind is prominent (see below) is 
very close to the cluster center), then the IMBH mass 
limit is correspondingly uncertain.

\subsection{Millisecond Pulsars: 47Tuc-W is not alone?}
The initial \chandra~ observation of 47Tuc showed 
the MSPs to be predominantly soft thermal 
sources (GHE01a), with 9 of the 15 then located by radio 
timing detected and several others plausibly detected. Detailed 
initial studies (Grindlay et al 2002) 
of the MSP \Lx vs. \Edot relation showed a 
significantly flatter dependence (\Lx $\propto$ \Edot$^{0.5}$) than 
the linear relation found for the predominantly magnetospheric 
emission from more luminous pulsars. This work has been extended 
with an attempt to separate the thermal vs. non-thermal MSP 
X-ray luminosities vs. \Edot (Grindlay 2005) using data from 
the 2002 deep dataset on 47Tuc. A more detailed study is nearing 
completion (Bogdanov et al 2005, in preparation; hereafter 
BGH05). 

A key result, first realized at this Cefalu meeting, is that the 
eclipsing MSP-W, first located from the HST discovery of its 
optical companion (Edmonds et al 2002), is  remarkably similar 
to the quiescent low mass X-ray binary (qLMXB) and first-discovered 
accreting millisecond pulsar, J1808-3658. The deep \chandra~ data 
showed (Bogdanov, Grindlay and van den Berg 2005; 
hereafter BGvB05) that its X-ray lightcurve shows broad eclipses 
of its hard flux but  no evidence for the sharp and total eclipse 
expected for the soft thermal component from the NS (Figure 2a).  
This can be explained by the system geometry shown in Figure 2b: 
the MSP wind produces a standing shock at (or near) the L1 point, 
where mass from the secondary is overflowing the Roche lobe, and 
non-thermal (synchrotron) emission from this shock is then eclipsed 
by the secondary at binary phases $\phi$ \about0.4 - 0.6. The 
longer-rise egress is due to the longer visibility 
of the swept-back emission region.

\begin{figure}
\includegraphics[height=.26\textheight,clip]{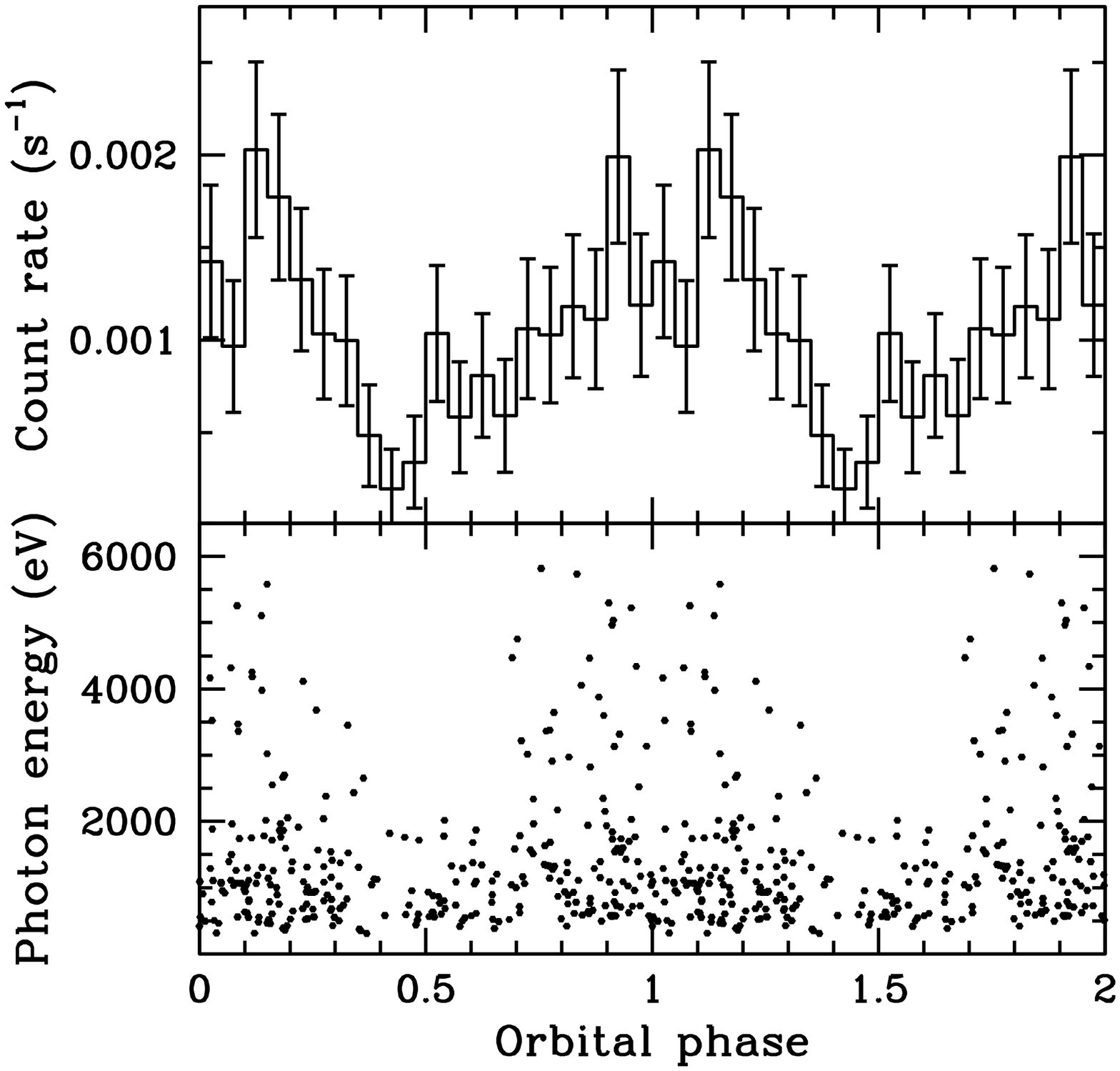}
\includegraphics[height=.26\textheight,clip]{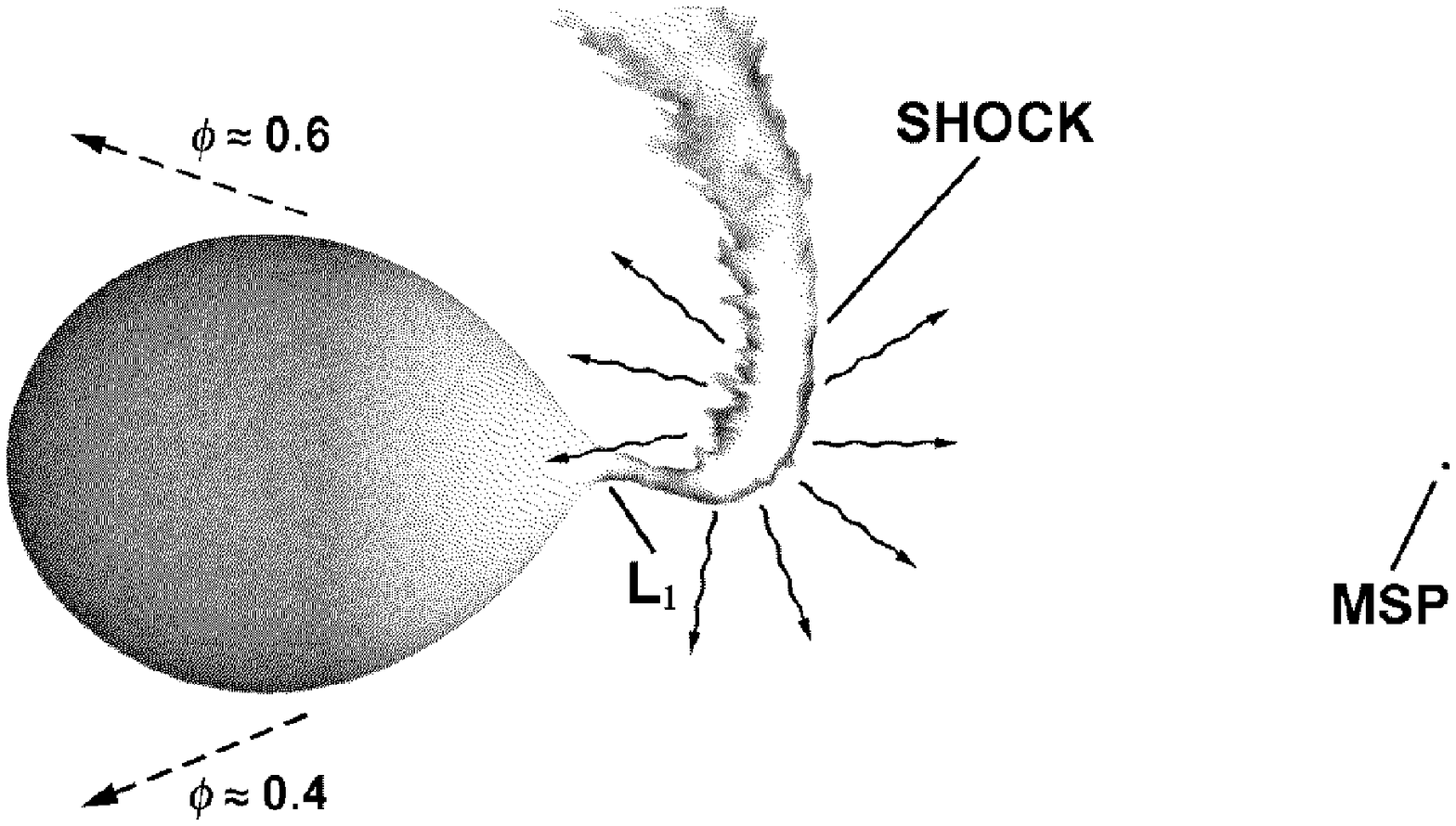}
\caption{{\it Left:} a) Light curves of MSP-W showing total 
counts and counts vs. energy folded on the 3.2h binary 
period. {\it Right:} b) Sketch of MSP-W to explain eclipse 
geometry (from BGvB05).}
\end{figure}

A complete study of the full sample of 19 MSPs with precise 
locations, now including the new timing positions for 
MSPs -R and -Y (Freire et al 2005, in preparation), is presented by 
BGH05 and shows that at least 3 (-W, -J and -O) 
of the 19 MSPs show significant emission above 2 keV. Certainly 
for -W, and probably also for the eclipsing systems -J and -O,  
this is non-thermal (PL) emission from 
shocked gas excited by the MSP wind. This component for 
MSP X-ray emission was found (GCH02) to be dominant in the first (and 
still only) MSP in NGC 6397, N6397-A, which, like that for 
47Tuc-W, is probably filling its Roche lobe but not able to accrete. 
Both have ``red straggler'' (to red of cluster main sequence) 
companions, which may indicate mass loss from a sub-giant (N6397-A) 
or a puffed up envelope for a main sequence star (47Tuc-W). This may  
be due to the dynamical heating of the companions in their re-exchange. 
Two of the CVs in NGC 6397 may also have ``puffed up'' 
companions given their binary periods vs. inferred companion masses  
(Taylor et al 2005; hereafter TGE05).

\subsection{CVs vs. ABs: Distinct spectral variability}
Another surprise revealed by the initial 47Tuc observation 
were the luminosities and spectral hardness of the active binaries. 
This is extended in the 2002 observation, with ABs making up 
the largest total population of sources with (likely) 
optical IDs: extrapolating from the HST coverage used by 
Edmonds et al (2003) for IDs and extended by HGE05 for 
initial IDs from 2002 data, the total AB source detections 
may be as high as 178 vs. 113 for the CVs. Whereas the AB vs. 
CV X-ray luminosity functions (XLFs) cross at \Lx \about10$^{30.5}$ 
\lcgs, with CVs definitely dominant (16:3) 
for \Lx \gsim 10$^{31}$ \lcgs, 
distinguishing these two very different binary populations below 
this divide is increasingly difficult given the similarly hard 
spectra (in many cases) and even short-timescale variablity. The 
latter is revealed by the very similar flare like behaviour of 
the CV W51 vs. the AB W47, as shown in Fig. 6 of HGE05: both 
show factor of \about20 increases in count rate (0.5-6keV) with 
rise times \about30min and durations \about2-3h. This is perhaps 
``typical'' for very large flares on BY Dra or RS CVn ABs, but 
is unprecedented for a CV. It is not clear if this is an 
accretion instability or how rare such giant CV flares are: W51 
had two other smaller flares, each with a 
\about6\X increase and  similar 
timescales, in the third of the four 
65ksec obserations in the 2002 data), and only one other CV (W2) 
had two comparable (\about5\X increase) closely-spaced 
flares out of the lightcurves examined for the 22 optically 
identified CVs for each of the 4 \X ~65ksec observations.

\begin{figure}
\includegraphics[height=.7\textheight,angle=90.,clip]
{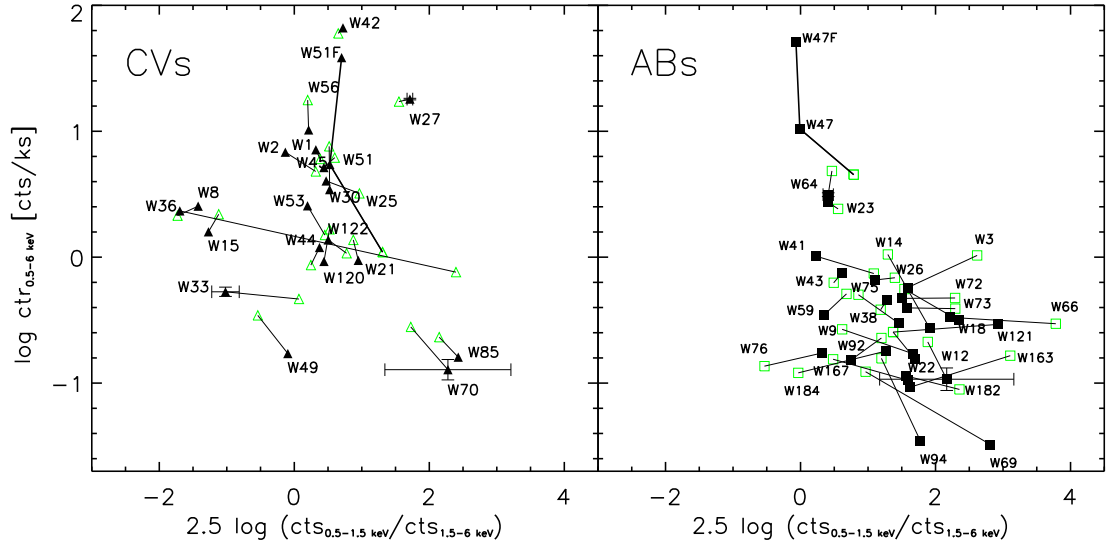}
\caption{X-ray color magnitude diagrams (XCMDs) of {\it Left:} a) 
CVs and  {\it Right:} b) ABs in 47Tuc. The 2000 (-I) data (open, 
green) are connected to the 2002 (-S) data (filled, black) 
for each source; the large flares for W47 and W51 
({\it not} removed from the total emission) 
are connected as a 3rd point (``F'').}
\end{figure}

The flare spectra, however, are 
very different and may point the way to distinguish ABs from CVs. 
The AB flare is very hard (vs. the quiescent spectrum), with a 
Bremsstrahlung fit yielding kT = 169 $\pm$29 keV and absorption 
column \nh = 7 \X 10$^{20}$ \cmsq vs. a quiescent spectrum (HGE05) 
with MEKAL fit of kT = 9 $\pm$1.3 keV and \nh = 11.8 $\pm$1.2 \X 
10$^{20}$ \cmsq. In contrast, the remarkable flare (or is it 
``super-blobby'' accretion?) from the CV W51 
is fit with (Brems) kT = 3.5  $\pm$1 keV 
and \nh = 2 \X 10$^{20}$ \cmsq, 
or {\it softer} than the quiescent emission (HGE05) with (MEKAL) 
kT = 5.7 $\pm$0.8 keV and \nh = 4.8 $\pm$1.1 \X 10$^{20}$ \cmsq.  
Harder spectra are typical of stellar (or solar) flares, where 
non-thermal processes dominate, whereas an accretion instability 
and higher \mdot for a CV would be expected to be more optically 
thick and thus softer. 

Spectral variability differences 
for the identified CVs vs. ABs in 47Tuc 
should be evident in XCMDs  
for the 2000 vs. 2002 observations (Figure 3). 
The ACIS-I counts have 
been transformed to those expected for ACIS-S for the nominal 
spectra for quiescent emission, taken to be (Brems) kT = 10 keV 
(CVs) vs. kT = 1 keV (ABs). Comparing sources 
with at least one of the two measurments 
\gsim1ct(0.5-8keV)/ksec for minimal error bars, the 18 CVs 
show 8:6:4 with positive:negative:uncertain slopes vs. 
2:4:0 for the 6 ABs, so the differences are only suggestive.  

\section{vs.~ NGC 6397}

\begin{figure}
\includegraphics[height=.2\textheight,clip]{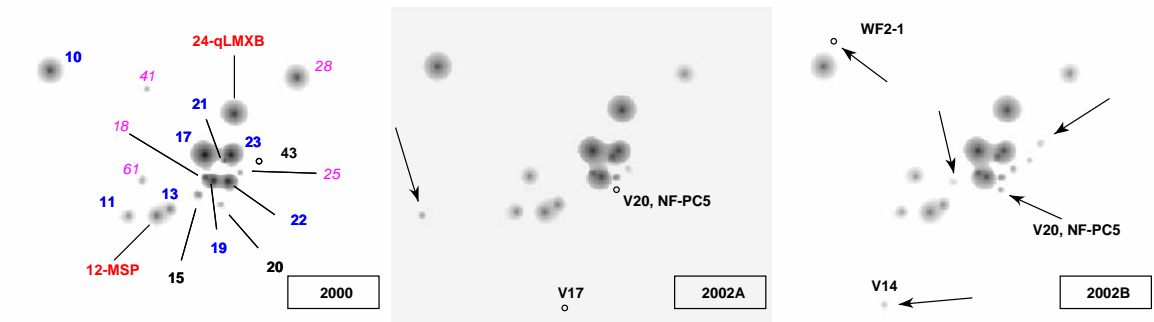}
\caption{\chandra~ images of central region of NGC 6397 from 
ACIS-I (left; see GHE01b) and -S (middle and right; see GvBB05). 
Sources are numbered as in GHE01b and new sources are 
marked. CV source numbers are blue, ABs are black, the MSP and qLMXB are red, 
and unidentified source {\it numbers} are magenta.  Scale: CVs U17 and U23 
are 10.0\arcsec apart. The cluster center is about 1\arcsec 
from U19/CV2.}
\end{figure}

NGC 6397 is the perfect ``foil'' for 47Tuc: it is core 
collapsed, with a power law cusp in its core flattening 
to a core radius with the latest estimate from HST  
(for main sequence stars) as 
\rc = 4.4 $\pm3.2$\arcsec (TGE05)
rather than the ``perfect'' King model with \rc =24\arcsec 
for 47Tuc. Given the factor of \about2 closer distance of NGG 6397 
(2.3kpc), the isothermal core is \about8\X 
smaller in radius than that for 47Tuc and thus for the quoted  
stellar luminosity density (log $\rho_L$ \about5.68 vs. 4.81 
for 47Tuc) has a core mass (and IXB factory) 
some \about400\X smaller than that of 47 Tuc. 
It is metal-poor ([Fe/H] = -2.0 vs. -0.7 for 47Tuc) and 
has absolute magnitude -6.63 vs. -9.42 for 47Tuc, suggesting a 
mass ratio of 13 for constant M/L. Thus it provides a contrast in 
its dynamical history, metallicity and mass. 

Our initial 50 ksec observation of NGC 6397 with ACIS-I in July 2000 
(Grindlay et al 2001b; hereafter GHE01b) was followed up with 
a comparable exposure (2 \X ~25 ksec) with ACIS-S 
on May 13 and 15, 2002. 
Details are reported in Grindlay et al (2005; hereafter GvBB05);  
highlights for comparison with 47Tuc are reported here. The 
Wavdetect images of the ACIS-I vs. -S exposures are shown in Figure 4. 
In the core region shown, sources U15, U20 and U41 
are detected only in the ACIS-I observation  
and 6 new sources (marked with arrows) are detected in one or both 
of the -S exposures. The Vxx designations mark identifications 
with variable stars in the cluster reported by Kaluzny and 
Thompson (2003) and references therein. Open circles 
mark lower-threshold detections with one either an AB (V20) 
or possibly a ``non-flickerer'' (NF-PC5; a probable He WD, 
see Taylor et al 2001) -- both are within 
the 95\% confidence radius (\about0.5\arcsec).  
Additional source details and 
identifications with HST stars are given in GvBB05 and TGE05.
Additional sources of note are outside the field shown (e.g. U60, 
recently identified with HST by TGE05 as the 9th CV, is along 
the U24 - U12 line to the SE). Source U28, tentatively 
classified as a CV by GHE01b from its hard spectrum, is 
in fact identified by Cool et al (in preparation) as 
a background edge-on Seyfert galaxy(!) so that the cluster CV total 
from HST/WFPC2 identifications (TGE05) remains at 9.

Clearly many sources are highly variable. U22 = CV5 decreased 
its X-ray luminosity by a factor of \about10 and the ABs show 
large variations. The XCMD for the ACIS-I vs. -S spectral variability 
of the CVs and HST-identified (TGE05) CVs is shown in Figure 5 
for comparison with the 47Tuc plot (Fig. 3). Although there are 
fewer objects, the CVs now all show brighter-softer 
variations and the ABs (with larger errors) are again 
brighter-harder. All four CVs (CV1/U23, CV2/U19, CV4/U21 
and CV6/U10) with binary periods discovered  
with HST (TGE05) show significant modulation in the 
combined \chandra~ data; details are discused by GvBB05. 

\begin{figure}
\includegraphics[height=.7\textheight,angle=90.,clip]
{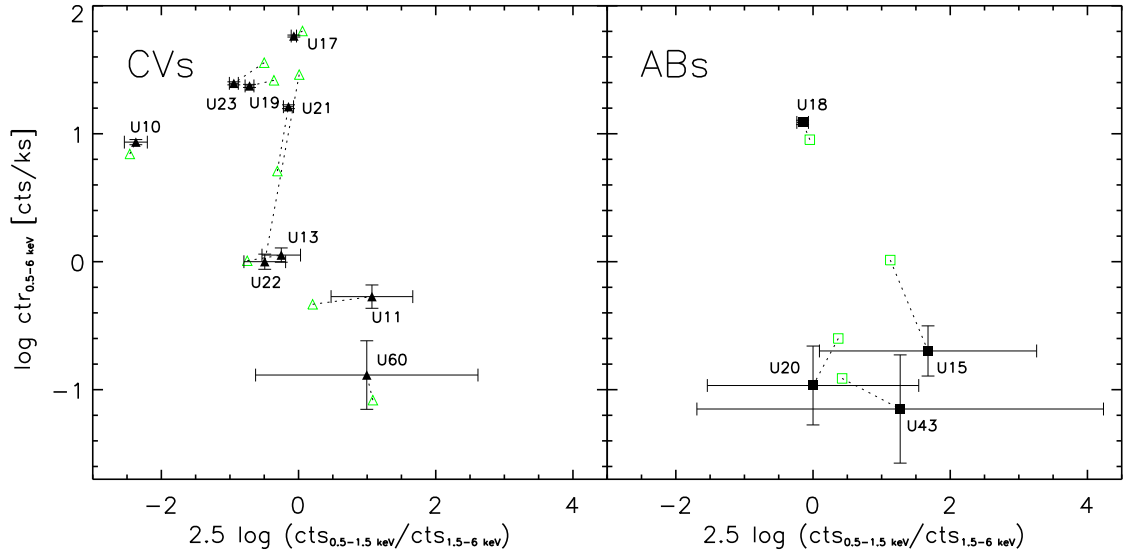}
\caption{X-ray color magnitude diagrams (XCMDs) of {\it Left:} a) 
confirmed CVs and  {\it Right:} b) candidate ABs in NGC 6397. 
Colors and labels for the 2000 vs. 2002 data are as in Fig. 3.}
\end{figure}

\section{Do IXBs Trace Cluster Formation and Evolution?}
The IXB populations and spatial distributions in both 
clusters can be compared for constraints on compact 
object populations and IXB formation/evolution. First, since 
the CVs and qLMXB/MSP systems are both significantly 
over-produced, by factors of \gsim10, in these clusters vs. 
the field populations (see GvBB05), both are produced in 
exchange encounters (primarily) with primordial binaries or, 
particularly during core collapse for NGC 6397, by tidal capture. 
NSs are more concentrated in the core and thus favored in 
IXB production, and lower mass WDs may be more 
easily expelled than NSs in the 
core collapse that has occurred in NGC 6397. 
Thus if both clusters have the same initial mass functions (IMFs) 
and thus ultimately ratios of white dwarfs (WDs) 
to neutron stars (NSs), the ratios of IXBs containing each 
should be similar in both clusters, with perhaps more NS-IXBs 
expected in NGC 6397. 

\subsection{NSs vs. WDs in IXBs: cluster IMF?}
In fact, as we originally 
suspected (GHE01b), the relative numbers of WDs vs. NSs 
locked up in IXBs in 47Tuc vs. NGC 6397 are surely not the same.  
In Table 1 we summarize the observed IXBs for both clusters and 
derive the ratio of expected IXBs containing WDs vs. NSs 
{\it normalized} by the 
ratio of relative collision number, or rate of IXB production, 
per unit cluster mass. The scaling for collision number, 
in a cluster core with density $\rho_c$ and core radius \rc is  
$\Gamma_c \propto \rho_c^{1.5}$ \rc$^2$, is taken from Verbunt (2003) 
and Heinke et al (2003). The bottom line is that 
the NS/WD ratio in IXB systems appears to be 
enhanced by a factor of \about6 in 47Tuc vs. NGC 6397. It is 
conceivable that the core collapse in NGC 6397 would 
favor production of WD-IXBs by tidal capture simply 
because WDs are so much more numerous than NSs in any cluster. 
However binary ``burning'' to halt core collapse presumably 
produces a net loss of binaries in the core (and indeed the ABs 
appear in Figure 4 to be relatively deficient near the cluster 
center and have a core radius measured (TGE05) to be significantly 
larger (\rc = 9.3$\pm3.5$\arcsec) than the CVs 
(\rc = 1.0$\pm3.5$/arcsec). If the NS/WD ratio difference is 
due to CV production in core collapse, then the youngest cluster 
CVs should be in the central core. It is interesting that 
the three optically faintest CVs (7, 8 and 9 = U11, U13, and U60), 
all with absolute magnitudes Mv \about 11.5 - 12 and companion 
masses thus \about0.1\Msun~ (TGE05) are indeed farther out from 
the cluster center as expected if CVs 1-6 were more recently created.   

If CV (vs. LMXB) production is not favored in core collapse, then the 
lower NS/WD ratio in NGC 6397 points to the underlying NS population 
being deficient. This could be (partly) due to the lower escape 
velocity for its reduced cluster mass, but the present mass may be 
greatly reduced by tidal stripping and so the initial masses 
are obviously uncertain. More likely, the large metallicity 
difference suggests a much flatter IMF for 47Tuc and thus NS 
initial production, as originally suggested by Grindlay (1987) 
for the luminous galactic globular cluster LMXBs and as now 
suggested for the pronounced metallicity dependence of 
extragalactic cluster LMXBs (e.g., Jordan et al 2004).

\begin{table}
\begin{tabular}{lrrr}
\hline
  & \tablehead{1}{r}{b}{Source Type}
  & \tablehead{1}{r}{b}{47Tuc}
  & \tablehead{1}{r}{b}{NGC 6397}   \\
\hline
Observed: & qLMXBs (NSs) & 5        & 1 \\
          & MSPs (NSs)   & \about30\tablenote{MSP and CV 
numbers for both clusters are estimated totals} & \about2 \\
          & CVs (WDs)     & \about30 & \about12 \\
%          & ABs (MSs)     & \about60 & \about15 \\
          & NSs/WDs in IXBs & \about1 & \about0.25 \\
Derived:  & $\Gamma_c$ (rel. coll. rate) & \about3  & \about0.3 \\
          & M$_{GC}$ (rel. total mass) & \about13 & \about1 \\
          & (NS/WD) / ($\Gamma_c$/M$_{GC}$) & \about5 & \about0.8 \\ 
%          & ABs / ($\Gamma_c$/M$_{GC}$) & \about260  & \about50 \\
\hline
\end{tabular}
\caption{Comparison IXB counts: 47Tuc vs. NGC 6397}
\label{tab:a}
\end{table}

\subsection{Interacting binaries vs. cluster rotation?}
As noted in GHE01b and as now even more apparent in Figure 4, the 
IXBs in NGC 6397 appear possibly anisotropic: at radii \lsim1\arcmin~ 
from the cluster center they are predominantly scattered along a NW-SE 
``line'' which is matched by the ``line''  of 5-6 luminous blue 
stragglers in the core. With more sources (mostly ABs) now detected 
in the cluster (Figure 4), this trend is strengthened (e.g. 5 of the 
6 ``new'' sources are roughly along this axis). Very similar 
cluster core flattening of the \chandra~ source distributions are 
apparent for the globulars NGC 6440 and NGC 6266 presented by 
Pooley et al (2002). 47Tuc also shows an apparent anisotropy of 
its NS-IXBs (MSPs and qLMXBs), as shown by Grindlay (2005) who 
examined correlations with the cluster proper motion.

A better (or more plausible) possible interpretation, if any 
of these anisotropies are significant (and simulations are 
planned) is to consider cluster rotation. Kim, Lee and Spurzem (2004) 
have found that core collapse and mass segregation are enhanced 
by cluster rotation. In Figure 6 the NS-IXBs for 47Tuc and 
the complete IXB sample for NGC 6397 are shown with the cluster 
rotation equator directions ($\pm2\sigma$) from the rotation 
measurements of Gephardt et al (1995) marked. For 47Tuc, the bulk 
of the MSPs and qLMXBs are within the $\pm2\sigma$ region around 
the rotational equator, and for NGC 6397 the flattened core 
region is at \about2.5$\sigma$ from the rotation equator 
line (but also, given the large uncertainties, \about2.5$\sigma$ 
from the rotation axis!). The position angle (PA) of the 
rotation velocity equator for NGC 6397 is 
very uncertain and only measured in integrated light; for 47Tuc 
the PA is in excellent agreement for both 
integrated light and individual stellar velocities. Large stellar 
velocity samples are needed to measure the stellar rotation for 
NGC 6397 (and NGC 6440 and NGC 6266) to test these  
alignments. 

\begin{figure}
\includegraphics[height=.3\textheight,angle=0.,clip]
{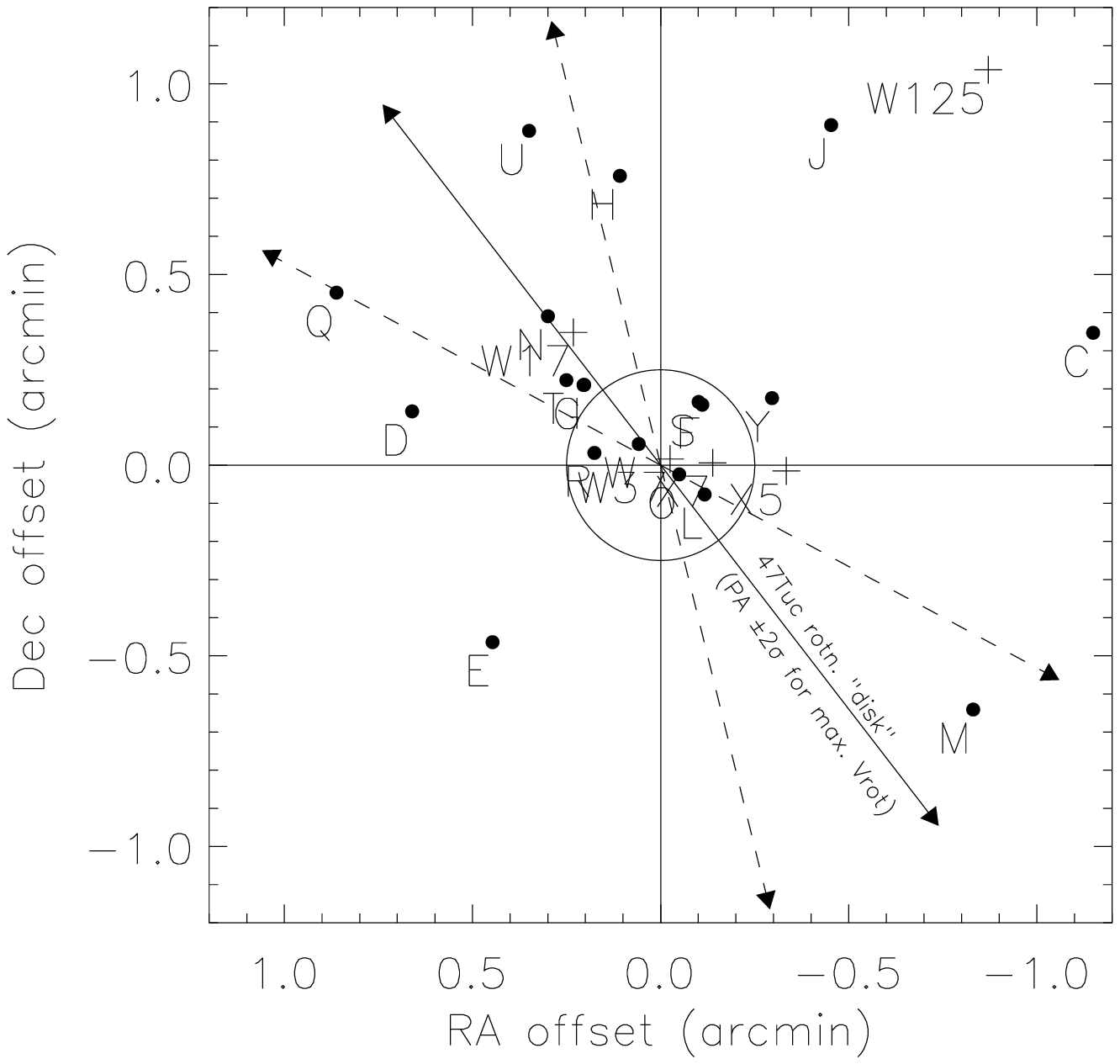}
\includegraphics[height=.3\textheight,angle=0.,clip]
{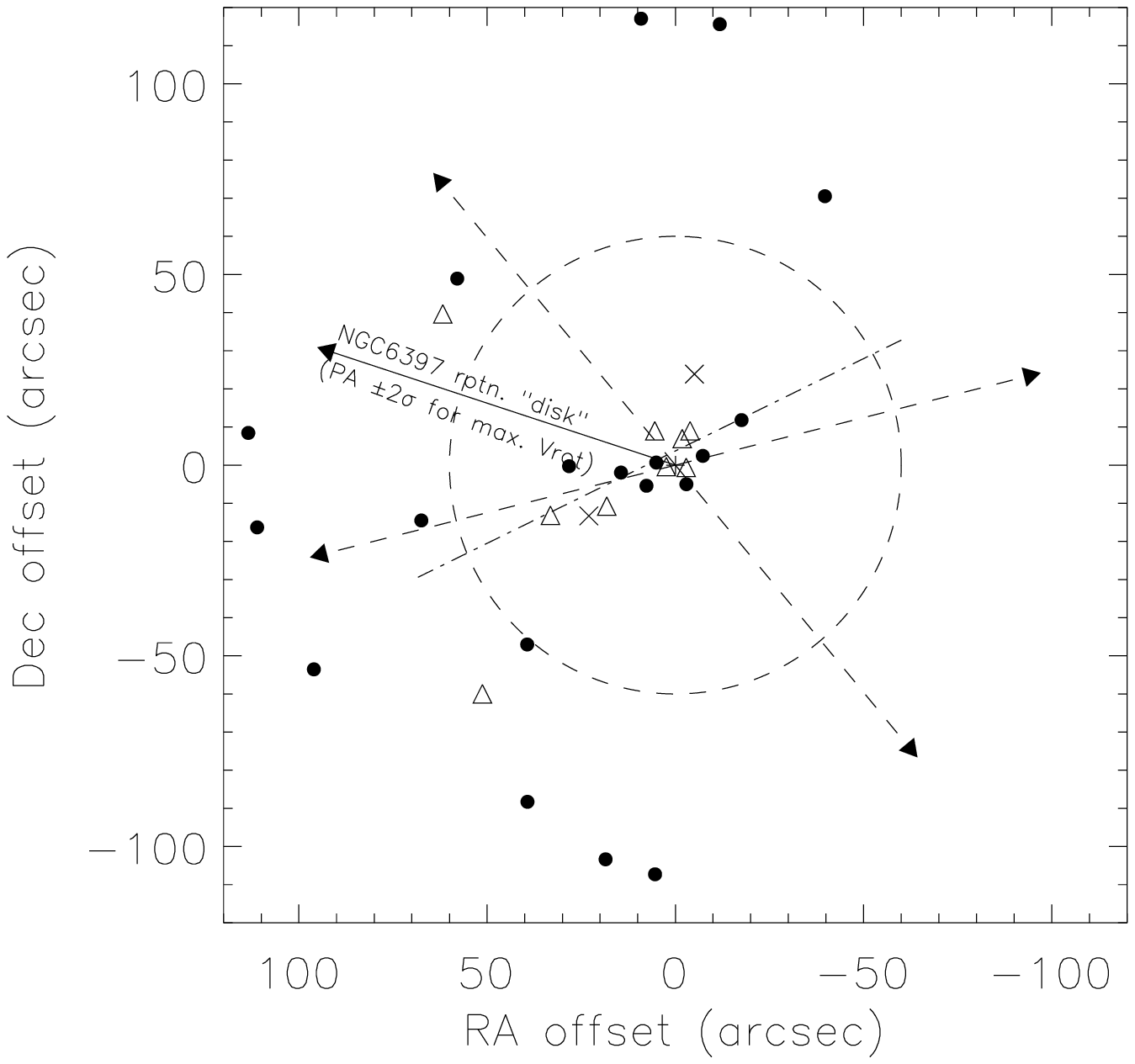}
\caption{IXB sources locations vs. cluster rotation equator for 
{\it Left:} a) MSPs (filled dots) and qLMXBs (+) in 47Tuc; circle 
marks optical core radius, 
and  {\it Right:} b) all sources in NGC 6397: triangles=CVs, X=MSP 
and qLMXB, and filled dots= ABs and unIDs. The source ``equatorial'' 
plane is fit with the dot-dashed line for sources inside the circle;  
exterior sources are increasingly background objects.}
\end{figure}

For 47Tuc, the implication of alignment of the NS-IXBs could be 
that the NSs, as the oldest objects in the cluster and which may 
have formed in the proto-cluster disk, have retained their mean 
angular orbital momentum despite having acquired companions in exchange 
collisions. Angular momentum alignment would remain fixed in 
the cluster frame, which was not necessarily the case for 
anisotropies induced by the cluster proper 
motion (Grindlay 2005). For NGC 6397, the most natural 
expectation is that rotational flattening during core collapse 
has scattered IXBs preferentially in the equatorial plane. 

\section{Conclusions}
Interacting binaries in globular clusters are allowing new domains of 
stellar and binary evolution to be studied: from clues to the
formation of the very first massive stars (and NSs), to the oldest 
and least luminous CVs to the extremes of stellar binaries. They 
point the way to new dynamical phenomena, including re-re-cycling and 
alignment processes with cluster angular momentum. High resolution  
\chandra~ imaging has provided the key for new understanding and 
new questions.

%%%%%%%%%%%%%%%%%%%%%%%%%%%%%%%%%%%%%%%%%%%%%%%%
%% BACKMATTER
%%%%%%%%%%%%%%%%%%%%%%%%%%%%%%%%%%%%%%%%%%%%%%%%

\begin{theacknowledgments}
I thank Maureen van den Berg, Slavko Bogdanov, and Craig Heinke 
for their many key contributions to the analysis summarized 
here. This work was supported in part by \chandra~ grant 
GO2-3059A and HST grant GO-0944.
\end{theacknowledgments}


%\endinput

\end{document}